\documentclass[12pt,reqno]{amsart}
\usepackage{graphicx}
\usepackage{amscd}
\usepackage{amsmath}
\usepackage{epsfig}
\usepackage{amsfonts}
\usepackage{amssymb}

\providecommand{\U}[1]{\protect\rule{.1in}{.1in}}
\providecommand{\U}[1]{\protect\rule{.1in}{.1in}}
\allowdisplaybreaks

\theoremstyle{plain}

\numberwithin{equation}{section}

\input{tcilatex}

\begin{document}
\title[The Logistic Model]{Logistic Models with Time-Dependent Coefficients
and Some of Their Applications}
\author{Raquel M. Lopez}
\address{Mathematical, Computational and Modeling Sciences Center, Arizona
State University, Tempe, AZ 85287--1904, U.S.A.}
\email{rlopez14@asu.edu}
\author{Benjamin R.~Morin}
\address{Mathematical, Computational and Modeling Sciences Center, Arizona
State University, Tempe, AZ 85287--1904, U.S.A.}
\email{brmorin@asu.edu}
\author{Sergei K. Suslov}
\address{Mathematical, Computational and Modeling Sciences Center, Arizona
State University, Tempe, AZ 85287--1904, U.S.A.}
\email{sks@asu.edu}
\urladdr{http://hahn.la.asu.edu/\symbol{126}suslov/index.html}
\date{\today }
\dedicatory{Dedicated to Doron Zeilberger on his 60th birthday.}

\begin{abstract}
We discuss explicit solutions of the logistic model with variable
parameters. Classical data on the sunflower seeds growth are revisited as a
simple application of the logistic model with periodic coefficients. Some
applications to related biological systems are briefly reviewed.
\end{abstract}

\maketitle

\section{Introduction}

We consider explicit solutions of logistic equations with time-dependent
parameters. Particular attention is paid to the case of periodic
coefficients which is especially important for many biological problems due
to a natural periodicity of the Earth rotations. Models of this type can be
used to mimic a population's response to seasonal fluctuations in its
environment or to mimic a population with several discrete life-cycle stages 
\cite{Boyce:Daley80}, \cite{Clark:Gross90}, \cite{Coleman80} \cite%
{Rodriguez88}. As an example, sunflower plant growth data from the classical
paper \cite{Reed:Holland1919} are reviewed and analyzed from the viewpoint
of the day-night periodicity. We dedicate this paper to Professor Doron
Zeilberger on his 60th birthday --- his nontraditional way of thinking has
motivated one of the authors (S.~K.~S.) to review this classical topic from
a modern perspective.

\section{Solution of the Logistic Model with Time-Dependent Coefficients}

The nonlinear equation for the logistic, or Verhulst's, model \cite%
{Brau:Cast}, \cite{Clark:Gross90}, \cite{Coleman80}, \cite{Verh1838}, \cite%
{Verh1841}, \cite{Verh1845}, \cite{Verh1847}:%
\begin{equation}
\frac{dx}{dt}=\alpha \left( t\right) x\left( 1-\frac{x}{\beta \left(
t\right) }\right)  \label{LogModEq}
\end{equation}%
with time-dependent coefficients $\alpha \left( t\right) $ and $\beta \left(
t\right) $ can be reduced to the linear equation 
\begin{equation}
z^{\prime }+\alpha \left( t\right) z=\frac{\alpha \left( t\right) }{\beta
\left( t\right) }  \label{LinEq}
\end{equation}%
by the standard substitution $z=1/x.$ Integration of (\ref{LinEq}) with the
help of an integrating factor,%
\begin{equation}
\frac{d}{dt}\left( z\exp \left( \int_{t_{0}}^{t}\alpha \left( s\right) \
ds\right) \right) =\frac{\alpha \left( t\right) }{\beta \left( t\right) }%
\exp \left( \int_{t_{0}}^{t}\alpha \left( s\right) \ ds\right) ,
\label{IntFact}
\end{equation}%
results in the explicit solution of the logistic model (in terms of a
weighted harmonic mean): 
\begin{equation}
\frac{1}{x\left( t\right) }=e^{-\int_{t_{0}}^{t}\alpha \left( s\right) \
ds}\left( \int_{t_{0}}^{t}\frac{\alpha \left( s\right) }{\beta \left(
s\right) }e^{\int_{t_{0}}^{s}\alpha \left( v\right) \ dv}\ ds+\frac{1}{%
x\left( t_{0}\right) }\right)  \label{LogModSol}
\end{equation}%
corresponding to the initial condition $x\left( t_{0}\right) =x_{0}.$ A
traditional expression is given by%
\begin{equation}
x\left( t\right) =\frac{x_{0}\exp \left( \dint_{t_{0}}^{t}\alpha \left(
s\right) \ ds\right) }{1+x_{0}\dint_{t_{0}}^{t}\dfrac{\alpha \left( s\right) 
}{\beta \left( s\right) }\exp \left( \dint_{t_{0}}^{s}\alpha \left( v\right)
\ dv\right) \ ds}  \label{LogModSolTrad}
\end{equation}%
(see also \cite{Boyce:Daley80}, \cite{Brau:Cast}, \cite{Clark:Gross90}, \cite%
{Coleman80}, \cite{Laksh03}, \cite{Laksh05}, \cite{Leach:Andriop04}, \cite%
{Tang:Cheke:Xiao06} and references therein). Here, we would like to consider
the most general functions $\alpha \left( t\right) $ and $\beta \left(
t\right) $ such that all integrals exist and%
\begin{equation}
\frac{d}{dt}\int_{t_{0}}^{t}f\left( s\right) \ ds=f\left( t\right)
\label{FandThCalcs}
\end{equation}%
by a corresponding version of the fundamental theorem of calculus (see, for
example, \cite{Kolm:Fom}, \cite{McDonald:Weiss}).

The well-known solution with constant parameters $\alpha $ and $\beta $ can
be written as%
\begin{equation}
\frac{1}{x\left( t\right) }=\frac{1}{\beta }+\left( \frac{1}{x\left(
t_{0}\right) }-\frac{1}{\beta }\right) e^{-\alpha \left( t-t_{0}\right) },
\label{ConsLogSol}
\end{equation}%
or%
\begin{equation}
x\left( t\right) =\frac{\beta x\left( t_{0}\right) }{x\left( t_{0}\right)
+\left( \beta -x\left( t_{0}\right) \right) e^{-\alpha \left( t-t_{0}\right)
}},  \label{ConstLogSol}
\end{equation}%
where $\beta =\lim_{t\rightarrow \infty }x\left( t\right) $ is the carrying
capacity.

\section{Logistic Models with Periodic Coefficients}

If variable coefficients $\alpha \left( t\right) $ and $\beta \left(
t\right) $ in the logistic equation (\ref{LogModEq}) are periodic functions
of time, namely,%
\begin{equation}
\alpha \left( t+T\right) =\alpha \left( t\right) ,\qquad \beta \left(
t+T\right) =\beta \left( t\right) ,  \label{Periodic}
\end{equation}%
an important question arises about the basic dynamics and long-term behavior
of a biological system under consideration. One may approach this matter in
the following mathematical setting. If the period is $T=t_{1}-t_{0},$ the
solution (\ref{LogModSol}) reads%
\begin{equation}
\frac{1}{x\left( t_{1}\right) }=a+\frac{b}{x\left( t_{0}\right) },
\label{DiscLogEq}
\end{equation}%
where parameters are given in terms of integrals over the period:%
\begin{equation}
a=b\int_{t_{0}}^{t_{1}}\frac{\alpha \left( t\right) }{\beta \left( t\right) }%
\exp \left( \int_{t_{0}}^{t}\alpha \left( s\right) \ ds\right) \ dt,\qquad
b=\exp \left( -\int_{t_{0}}^{t_{1}}\alpha \left( t\right) \ dt\right) .
\label{DisrLogPars}
\end{equation}%
Repeating this process indefinitely, we arrive at the following discrete map%
\begin{equation}
\frac{1}{x_{n+1}}=a+\frac{b}{x_{n}},  \label{DiscrMapLog}
\end{equation}%
where by the definition $x\left( t_{n}\right) =x_{n}.$ This recurrence
relation can be rewritten as follows%
\begin{equation}
x_{n+1}-x_{n}=\left( 1-b\right) x_{n+1}\left( 1-\frac{a}{1-b}x_{n}\right) ,
\label{DiscrEqLog}
\end{equation}%
which can be thought of as a difference analog of Verhulst's model (\ref%
{LogModEq}) with constant parameters. We refer to (\ref{DiscrMapLog}) as a
discrete logistic map (with a slight change of parameters it is also called
the Beverton--Holt difference equation \cite{Bev:Holt56}, \cite{Brau:Cast}, 
\cite{Tang:Cheke:Xiao06}).

But the first order nonhomogeneous difference equation of the form%
\begin{equation}
z_{n+1}=a+bz_{n}  \label{FirstDiffEq}
\end{equation}%
with constant coefficients $a$ and $b$ has the following explicit solutions 
\cite{Brau:Cast}, \cite{Clark:Gross90}:%
\begin{equation}
z_{n}=a\frac{1-b^{n}}{1-b}+b^{n}z_{0},\qquad \text{if\quad }b\neq 1,
\label{FirstDiffSol}
\end{equation}%
and%
\begin{equation}
z_{n}=an+z_{0},\qquad \text{if\quad }b=1,  \label{SecondDiffSol}
\end{equation}%
which can be verified by a direct substitution. Therefore the discrete
logistic map solution has the form%
\begin{equation}
\frac{1}{x_{n}}=\frac{a}{1-b}+\left( \frac{1}{x_{0}}-\frac{a}{1-b}\right)
b^{n}  \label{DiscrMapSols}
\end{equation}%
for any initial data $x_{0}.$ If $b<1,$ this sequence converges to the
limit, say $x_{\infty }:=\lim_{n\rightarrow \infty }x_{n},$ given by%
\begin{equation}
\frac{1}{x_{\infty }}=\frac{a}{1-b}  \label{Limit}
\end{equation}%
and our solution (\ref{DiscrMapSols}) takes a more convenient form%
\begin{equation}
\frac{1}{x_{n}}=\frac{1}{x_{\infty }}+\left( \frac{1}{x_{0}}-\frac{1}{%
x_{\infty }}\right) b^{n}.  \label{ExplicitSols}
\end{equation}%
Then%
\begin{equation}
\frac{x_{n}-x_{\infty }}{x_{n}}=\frac{x_{0}-x_{\infty }}{x_{0}}\ b^{n},
\label{ExplicitBounds}
\end{equation}%
which implies that $x_{n}>x_{\infty }$ for all $n,$ if $x_{0}>x_{\infty }$
and vice versa. Finally, from (\ref{ExplicitSols}) one gets%
\begin{equation}
-\frac{d}{dn}\left( \frac{1}{x_{n}}\right) =\frac{1}{x_{n}^{2}}\frac{dx_{n}}{%
dn}=\frac{x_{0}-x_{\infty }}{x_{0}x_{\infty }}\ b^{n}\ln b,  \label{nDeriv}
\end{equation}%
which reveals monotonicity properties of this bounded sequence in general.
Namely, our sequence is strictly increasing, when $x_{0}<$ $x_{\infty },$
and strictly decreasing otherwise.

Traditionally, a special question which arises in the theory of equations
with periodic coefficients is the existence of periodic solutions \cite%
{Arscott}, \cite{Mag:Win}. In the case of the logistic model under
consideration, when an explicit solution is available, one can only obtain a
continuous periodic solution by letting $x_{1}=x_{0}$ in our equation (\ref%
{DiscLogEq}). Thus%
\begin{equation}
\frac{1}{x_{0}}=a+\frac{b}{x_{0}},  \label{InitialPeriodic}
\end{equation}%
which corresponds to the special initial condition given by the expression (%
\ref{Limit}). According to (\ref{ExplicitSols}) the simple stability
property holds, if $b<1,$ solutions with all other initial data converge to
this periodic solution, or an attractor, as time goes to infinity with
precise monotonicity properties described above. By (\ref{ExplicitSols})%
\begin{equation}
\frac{1}{x_{n}}-\frac{1}{y_{n}}=\left( \frac{1}{x_{0}}-\frac{1}{y_{0}}%
\right) b^{n},  \label{Contructing}
\end{equation}%
which reveals the contracting map property for two different initial
conditions $x_{0}\neq y_{0}$ (see also \cite{Clark:Gross90} and \cite%
{Coleman80}).

\section{Two-Stage Compound Logistic Models}

In general, splitting a time interval $\left[ t_{0},t_{2}\right] $ as the
union of two successive subintervals, say $\left[ t_{0},t_{2}\right] =\left[
t_{0},t_{1}\right] \cup \left[ t_{1},t_{2}\right] ,$ one can write%
\begin{equation}
\frac{1}{x\left( t_{1}\right) }=a_{0}+\frac{b_{0}}{x\left( t_{0}\right) }%
,\qquad \frac{1}{x\left( t_{2}\right) }=a_{1}+\frac{b_{1}}{x\left(
t_{1}\right) },  \label{StepGen}
\end{equation}%
where the corresponding coefficients are given by our general expressions (%
\ref{DisrLogPars}). Then%
\begin{equation}
\frac{1}{x\left( t_{2}\right) }=a+\frac{b}{x\left( t_{0}\right) },
\label{StepGeneral}
\end{equation}%
where the `compound' coefficients are given by the following analog of the
addition formula%
\begin{equation}
a=a_{1}+a_{0}b_{1},\qquad b=b_{0}b_{1},  \label{composition}
\end{equation}%
which can also be derived directly from (\ref{DisrLogPars}). It is worth
noting the following group property%
\begin{equation}
\left( 1,\frac{1}{x\left( t_{0}\right) }\right) \left( 
\begin{array}{cc}
1 & a_{0} \\ 
0 & b_{0}%
\end{array}%
\right) =\left( 1,\frac{1}{x\left( t_{1}\right) }\right) ,\quad \left( 1,%
\frac{1}{x\left( t_{1}\right) }\right) \left( 
\begin{array}{cc}
1 & a_{1} \\ 
0 & b_{1}%
\end{array}%
\right) =\left( 1,\frac{1}{x\left( t_{2}\right) }\right)  \label{Vectors}
\end{equation}%
and%
\begin{equation}
\left( 
\begin{array}{cc}
1 & a_{0} \\ 
0 & b_{0}%
\end{array}%
\right) \left( 
\begin{array}{cc}
1 & a_{1} \\ 
0 & b_{1}%
\end{array}%
\right) =\left( 
\begin{array}{cc}
1 & a_{1}+a_{0}b_{1} \\ 
0 & b_{0}b_{1}%
\end{array}%
\right) .  \label{Matrix}
\end{equation}%
(Another applications of the group of upper triangular matrices can be found
in Refs.~\cite{Lop:Sus}, \cite{Ni:Su:Uv} and \cite{Vil}.)

With a slight change of notations, consider two step functions on the
interval $\left[ 0,t_{1}\right] :$%
\begin{equation}
\alpha \left( t\right) =\left\{ 
\begin{array}{cc}
\alpha _{0}, & \text{if}\ 0\leq t<t_{0} \\ 
\alpha _{1}, & \text{if}\ t_{0}\leq t<t_{1}%
\end{array}%
\right. \qquad \beta \left( t\right) =\left\{ 
\begin{array}{cc}
\beta _{0}, & \text{if}\ 0\leq t<t_{0} \\ 
\beta _{1}, & \text{if}\ t_{0}\leq t<t_{1}%
\end{array}%
\right.  \label{Step2}
\end{equation}%
and then define them by `periodicity' for all $t\geq 0$ with the period $%
T=t_{1}.$ We shall refer to this case as a two-stage compound logistic model
(see also \cite{Laksh03} for a special case). Evaluation of the integrals in
(\ref{DisrLogPars}) gives%
\begin{equation}
a=\frac{1-\tau _{1}}{\beta _{1}}+\frac{\left( 1-\tau _{0}\right) \tau _{1}}{%
\beta _{0}},\qquad b=\tau _{0}\tau _{1},  \label{2StepPars}
\end{equation}%
where by definition%
\begin{equation}
\tau _{0}=e^{-\alpha _{0}t_{0}},\qquad \tau _{1}=e^{-\alpha _{1}\left(
t_{1}-t_{0}\right) }.  \label{2StepTaus}
\end{equation}%
As a result%
\begin{equation}
x_{\infty }=\lim_{n\rightarrow \infty }x_{n}=\frac{\beta _{0}\beta
_{1}\left( 1-\tau _{0}\tau _{1}\right) }{\beta _{0}+\beta _{1}\tau
_{1}-\left( \beta _{0}+\beta _{1}\tau _{0}\right) \tau _{1}}
\label{2StepLimit}
\end{equation}%
provided $\tau _{0}\tau _{1}<1.$

Two special cases are of particular interest. If $\alpha _{1}=0,$ then $\tau
_{1}=1$ and%
\begin{equation}
a=\frac{1-\tau _{0}}{\beta _{0}},\qquad b=\tau _{0}=e^{-\alpha
_{0}t_{0}},\qquad x_{\infty }=\beta _{0}.  \label{2StepLimitZero}
\end{equation}%
This will be useful in section~8. Another limit, $\beta _{1}\rightarrow
\infty ,$ corresponds to a resetting mechanism discussed in \cite%
{Lop:Cam:Sus} (see also \cite{Tang:Cheke:Xiao06}).

\section{$n$-Stage Compound Logistic Model}

Results from the previous section can be easily generalized in the following
fashion. Condider a partition $t_{0}=s_{0}<s_{1}<...<s_{n}=t_{1}$ of a time
interval $\left[ t_{0},t_{1}\right] =\cup _{i=0}^{n-1}\left[ s_{i},s_{i+1}%
\right] .$ Then with the help of (\ref{composition})--(\ref{Matrix}):%
\begin{equation}
\frac{1}{x\left( s_{n}\right) }=a+\frac{b}{x\left( s_{0}\right) },
\label{ABNSol}
\end{equation}%
where%
\begin{equation}
a=\sum_{i=0}^{n-1}a_{i}\dprod_{k=i+1}^{n-1}b_{k},\qquad
b=\dprod_{i=0}^{n-1}b_{i},  \label{ABN}
\end{equation}%
or in the matrix form%
\begin{eqnarray}
\dprod_{i=0}^{n-1}\left( 
\begin{array}{cc}
1 & a_{i} \\ 
0 & b_{i}%
\end{array}%
\right) &=&\left( 
\begin{array}{cc}
1 & a_{0} \\ 
0 & b_{0}%
\end{array}%
\right) \left( 
\begin{array}{cc}
1 & a_{1} \\ 
0 & b_{1}%
\end{array}%
\right) \times ...\times \left( 
\begin{array}{cc}
1 & a_{n-1} \\ 
0 & b_{n-1}%
\end{array}%
\right)  \label{ProdMatrix} \\
&=&\left( 
\begin{array}{cc}
1 & \sum_{i=0}^{n-1}a_{i}\dprod_{k=i+1}^{n-1}b_{k} \\ 
0 & \dprod_{i=0}^{n-1}b_{i}%
\end{array}%
\right)  \notag
\end{eqnarray}%
by induction. It is worth noting that this multiplication property holds for
any numbers $\left\{ a_{i}\right\} _{i=0}^{n-1}$ and $\left\{ b_{i}\right\}
_{i=0}^{n-1},$ thus generating solution of the following difference equation%
\begin{equation}
\frac{1}{x_{i+1}}=a_{i}+\frac{b_{i}}{x_{i}},  \label{GenDiffLog}
\end{equation}%
which is obtained here with the help of the group property of upper
triangular matrices (see also \cite{Clark:Gross90} and \cite%
{Tang:Cheke:Xiao06}). In this paper, we take $x_{i}=x\left( s_{i}\right) $
and use definitions (\ref{DisrLogPars}), namely,%
\begin{equation}
a_{i}=b_{i}\int_{s_{i}}^{s_{i+1}}\frac{\alpha \left( t\right) }{\beta \left(
t\right) }\exp \left( \int_{s_{i}}^{t}\alpha \left( s\right) \ ds\right) \
dt,\qquad b_{i}=\exp \left( -\int_{s_{i}}^{s_{i+1}}\alpha \left( t\right) \
dt\right)  \label{ABParsI}
\end{equation}%
for the corresponding subintervals $\left[ s_{i},s_{i+1}\right] $ with $i=0,$
$1,$ $...,$ $n-1.$

For step functions $\alpha \left( t\right) =\alpha _{i}$ and $\beta \left(
t\right) =\beta _{i}$ on subintervals $s_{i}\leq t<s_{i+1}$ one gets%
\begin{equation}
a_{i}=\frac{1-\tau _{i}}{\beta _{i}},\qquad b_{i}=\tau _{i}=e^{-\alpha
_{i}\left( s_{i+1}-s_{i}\right) }\quad (i=0,1,...,n-1)  \label{TauI}
\end{equation}%
and%
\begin{equation}
a=\sum_{i=0}^{n-1}\frac{1-\tau _{i}}{\beta _{i}}\dprod_{k=i+1}^{n-1}\tau
_{k},\qquad b=\prod_{i=0}^{n-1}\tau _{i}.  \label{ABTotal}
\end{equation}%
Further details are left to the reader.

\section{An Extension}

The following generalization of the logistic model:%
\begin{equation}
\frac{dx}{dt}=\alpha \left( t\right) x\left( 1-\frac{x^{\theta }}{\beta
\left( t\right) }\right) ,\qquad \theta =\text{constant}  \label{theta}
\end{equation}%
has been considered in Refs.~\cite{Alm:Dow:Mor:Mur}, \cite{Ayalaetal73} and 
\cite{Brau:Cast}. This equation is integrable by the substitution $%
z=1/x^{\theta }$ and the corresponding extension of (\ref{LogModSol}) takes
the form%
\begin{equation}
\frac{1}{x^{\theta }\left( t\right) }=e^{-\theta \int_{t_{0}}^{t}\alpha
\left( s\right) \ ds}\left( \theta \int_{t_{0}}^{t}\frac{\alpha \left(
s\right) }{\beta \left( s\right) }e^{\theta \int_{t_{0}}^{s}\alpha \left(
v\right) \ dv}\ ds+\frac{1}{x^{\theta }\left( t_{0}\right) }\right)
\label{LogSolTheta}
\end{equation}%
with $x\rightarrow x^{\theta }$ and $\alpha \rightarrow \theta \alpha .$ The
case of periodic coefficients is analyzed in a similar fashion with the help
of this substitution. Further details are left to the reader.

\section{Logistic Models and Riccati's Equation}

Considering a more general model%
\begin{equation}
\frac{dx}{dt}=\alpha \left( t\right) x\left( 1-\frac{x}{\beta \left(
t\right) }\right) +\gamma \left( t\right) ,  \label{RiccatiEq}
\end{equation}%
where $\alpha \left( t\right) ,$ $\beta \left( t\right) $ and $\gamma \left(
t\right) $ are suitable real valued functions of time only, one obtains the
so-called Riccati equation, which can also be reduced to a linear second
order differential equation by a certain substitution (see, for example,
Refs.~\cite{Cor-Sot:Lop:Sua:Sus} and \cite{HilleODE} for more details). The
case of periodic coefficients will be discussed elsewhere.

\section{An Application: Logistic Growth of Sunflower Seeds Revisited}

Logistic models are well-known in autocatalysis, population dynamics,
mathematical epidemiology and theory of fishing (see, for example, \cite%
{Bev:Holt56}, \cite{Brau:Cast}, \cite{Clark:Gross90}, \cite{Coleman80}, \cite%
{Pearl:Read20}, \cite{Tang:Cheke:Xiao06} and references therein). Among
other biological applications, we utilize the periodic logistic models,
described above in detail, to sunflower plant growth data from the classical
paper \cite{Reed:Holland1919}, where the sunflower \ (helianthus) was chosen
because of the fact that it grows without producing branches and it was
thought that measurements of height and weight represented the growth of the
entire organism with a fair degree of accuracy. Original experimental
results are presented in the following table:

\begin{center}
Sunflower height versus growing days by Reed and Holland (1919)

\begin{tabular}{||l||l||l||l||}
\hline\hline
days $n$ & observed mean height $x_{n}$(cm) & constant $K$(this paper) & $%
\ln \left( \frac{x_{n}}{x_{\infty }-x_{n}}\right) $ \\ \hline\hline
$0$ & $10.00$ & $0.04\allowbreak 0593$ & $-3.1966$ \\ \hline\hline
$7$ & $17.93\pm 0.14$ & $0.04119$ & $-2.\allowbreak 5798$ \\ \hline\hline
$14$ & $36.36\pm 0.43$ & $0.03852$ & $-1.\allowbreak 7917$ \\ \hline\hline
$21$ & $67.76\pm 0.78$ & $0.03\allowbreak 3353$ & $-1.0137$ \\ \hline\hline
$28$ & $98.10\pm 1.38$ & $0.03\allowbreak 2672$ & $-0.46643$ \\ \hline\hline
$35$ & $131.00\pm 1.73$ & $0.03\allowbreak 2005$ & $0.05\allowbreak 8956$ \\ 
\hline\hline
$42$ & $169.50\pm 2.21$ & $0.03\allowbreak 8430$ & $0.6902$ \\ \hline\hline
$49$ & $205.50\pm 2.92$ & $0.04\allowbreak 2069$ & $1.4336$ \\ \hline\hline
$56$ & $228.30\pm 3.41$ & $0.04\allowbreak 3129$ & $2.\allowbreak 1649$ \\ 
\hline\hline
$63$ & $247.10\pm 3.80$ & $0.05\allowbreak 2904$ & $3.\allowbreak 5083$ \\ 
\hline\hline
$70$ & $250.50\pm 3.76$ & $0.05\allowbreak 0188$ & $4.1372$ \\ \hline\hline
$77$ & $253.80\pm 3.99$ & $0.059799$ & $5.\allowbreak 8932$ \\ \hline\hline
$84$ & $254.50\pm 3.89$ & not available & not available \\ \hline\hline
\end{tabular}
\end{center}

\noindent The following solution of the continuous Verhulst model, see (\ref%
{ConsLogSol})--(\ref{ConstLogSol}),%
\begin{equation}
\log \frac{x_{n}}{254.5-x_{n}}=K\left( n-t_{1}\right)  \label{Interp}
\end{equation}%
with $K=0.0421$ and $t_{1}=34.2$ was originally suggested in order to
interpolate these data \cite{Reed:Holland1919}. Here, the carring capacity
was chosen to be $\beta =254.5$ and the time at which the growth has run
half way to equilibrium; that is the time at which $x=\beta /2$ and the
maxinum growth occurs; was estimated as $t_{1}=34.2$ days. It is worth
noting that the average value of the constant $K,$ reevaluated here from our
corrected version of the third column (without $n=0),$ is given by $%
K=0.042205,$ which is slightly different from the value $K=0.0421$ obtained
in the original paper \cite{Reed:Holland1919} (this average becomes $%
K=0.04\allowbreak 2071\approx $ $0.0421$ if our $n=0$ data is included).

A simple assumption of the day-night periodicity of the plant growth results
in our solution (\ref{ExplicitSols}) of the discrete Verhulst map, which can
be rewritten for the numerical analysis purposes in a more convenient form:%
\begin{equation}
\ln \left( \frac{x_{n}}{x_{\infty }-x_{n}}\right) =\ln \left( \frac{x_{7}}{%
x_{\infty }-x_{7}}\right) +\left( n-7\right) \ln \left( \frac{1}{b}\right) .
\label{Data2}
\end{equation}%
Then we match the data available from the table on the seven day scale only
from $x_{7}=$ $17.93$ till $x_{84}=254.50\approx x_{\infty }.$ Corresponding
data from the last column have been first analyzed with the help of a least
square linear curve fitting on the entire available time interval $\left[
7,77\right] $. As a result%
\begin{eqnarray}
\ln \left( \frac{x_{n}}{254.50-x_{n}}\right)
&=&-3.612328764+0.112057018181818180n  \label{Data1} \\
&=&0.112057018181818180\left( n-32.\,\allowbreak 237\right) ,  \notag
\end{eqnarray}%
which gives the following modified values of parameters $t_{1}=32.\,%
\allowbreak 237$ and $K=0.112057018181818180$ $/\ln 10=0.04\allowbreak 8666$
in the original formula (\ref{Interp}). Unfortunately, this formula does not
match observed data better then the original expression in Ref.~\cite%
{Reed:Holland1919} (even when one uses a quadratic least square
approximation for the last column; two (or several)-stage logistic models
seem more approapriate). An elementary data analysis of the last column
shows that the best linear approximation occurs during the time interval $%
\left[ 7,49\right] ,$ when by the least square method: 
\begin{eqnarray}
\ln \left( \frac{x_{n}}{254.50-x_{n}}\right)
&=&-3.106504286+0.0922278367346938688n  \label{Data0} \\
&=&0.0922278367346938688\left( n-33.68293561\right)  \notag
\end{eqnarray}%
with the best values of parameters $t_{1}=33.68293561$ and $K=$ $%
0.04005404057$ in the formula (\ref{Interp}). Original data are approximated
well by sigmoid shape function; the revised results are presented in the
following graph (see Figire~1).

Finally, we would like to apply our periodic two-stage (day-night) compound
logistic model to helianthus growth. One may speculate that the sunflower
plant was growing about fourteen hours during an average California day from
May to August of 1919 and was not growing at all during nights. Then by (\ref%
{2StepLimitZero}):%
\begin{equation}
b=\exp \left( -\frac{7}{12}\alpha _{0}\right) =b_{0}^{7/12},\qquad
b_{0}=b^{12/7}.  \label{DayNightPars}
\end{equation}%
If the (sun)light would be provided for twenty four hours every day, say in
a greenhouse, expression (\ref{Interp}) should be modified as follows 
\begin{equation}
\log \frac{x_{n}}{254.5-x_{n}}=K_{0}\left( n-t_{0}\right)  \label{GreenHouse}
\end{equation}%
with $K_{0}=$ $\left( 12/7\right) K=\allowbreak 0.06\allowbreak 8664$ and $%
t_{0}=\left( 7/12\right) t_{1}=19.\allowbreak 648.$ The corresponding graph
is also presented on Figure~1.

%
\begin{figure}[htbp]
\centering \scalebox{.75}{\includegraphics{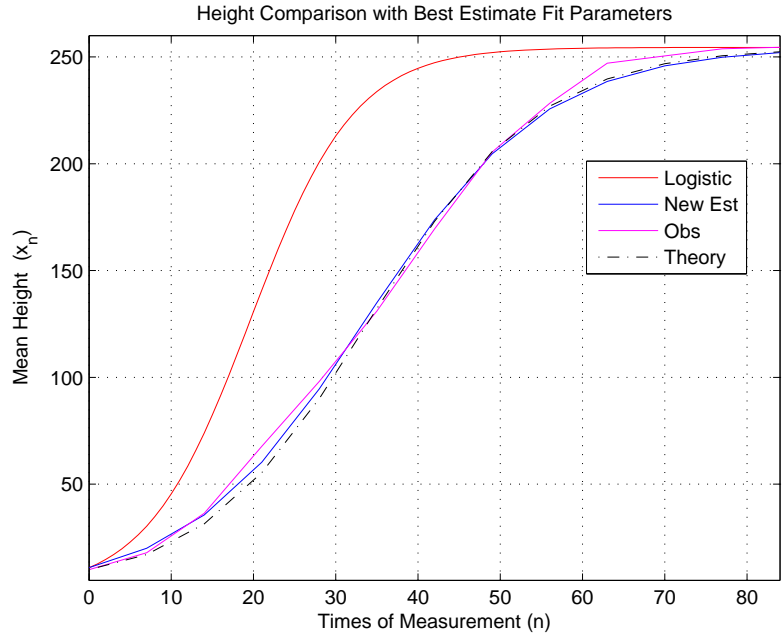}}
\caption{Comparison of observed and calculated values for the mean hight of
helianthus.}
\end{figure}

\noindent \textbf{Acknowledgments.\/} We thank Professor Carlos Castillo-Ch%
\'{a}vez, Professor Erika Camacho, Dr.~Carlos Castillo-Garsow, Professor Hal
Smith and Professor Doron Zeilberger for support, valuable discussions and
encouragement. This paper is written as a part of the summer 2010 program on
analysis of Mathematical and Theoretical Biology Institute (MTBI) and
Mathematical, Computational and Modeling Sciences Center (MCMSC) at Arizona
State University. The MTBI/SUMS Summer Research Program is supported by The
National Science Foundation (DMS-0502349), The National Security Agency
(DOD-H982300710096), The Sloan Foundation and Arizona State University.


\end{document}